\documentclass[twocolumn,superscriptaddress,aps,prl]{revtex4-2}

\usepackage{graphicx}
\usepackage{dcolumn}
\usepackage{bm}
\usepackage{amsmath}
\usepackage{amssymb}
\usepackage{siunitx}
\usepackage{upgreek}

\begin{document}

\title{Direct accessibility of the fundamental constants governing light-by-light scattering}

\author{Felix Karbstein}\email{f.karbstein@hi-jena.gsi.de}
\affiliation{Helmholtz-Institut Jena, Fr\"obelstieg 3, 07743 Jena, Germany}
\affiliation{GSI Helmholtzzentrum f\"ur Schwerionenforschung, Planckstra\ss e 1, 64291 Darmstadt, Germany}
\affiliation{Faculty of Physics and Astronomy, Friedrich-Schiller-Universit\"at Jena, 07743 Jena, Germany}
\author{Daniel Ullmann}
\affiliation{Faculty of Physics and Astronomy, Friedrich-Schiller-Universit\"at Jena, 07743 Jena, Germany}
\author{Elena A. Mosman}
\affiliation{Helmholtz-Institut Jena, Fr\"obelstieg 3, 07743 Jena, Germany}
\affiliation{Faculty of Physics and Astronomy, Friedrich-Schiller-Universit\"at Jena, 07743 Jena, Germany}
\author{Matt Zepf}\email{m.zepf@hi-jena.gsi.de}
\affiliation{Helmholtz-Institut Jena, Fr\"obelstieg 3, 07743 Jena, Germany}
\affiliation{GSI Helmholtzzentrum f\"ur Schwerionenforschung, Planckstra\ss e 1, 64291 Darmstadt, Germany}
\affiliation{Faculty of Physics and Astronomy, Friedrich-Schiller-Universit\"at Jena, 07743 Jena, Germany}

\date{\today}

\begin{abstract}
Quantum field theory predicts the vacuum to exhibit a non-linear response to strong electromagnetic fields. This fundamental tenet has remained experimentally challenging and is yet to be tested in the laboratory.
We present proof of concept and detailed theoretical analysis of an experimental setup for precision measurements of the quantum vacuum signal generated by the collision of a brilliant x-ray probe with a high-intensity pump laser.
The signal features components polarised parallel and perpendicularly to the incident x-ray probe. 
Our proof-of-concept measurements show that the background can be efficiently suppressed by many orders of magnitude which should not only facilitate a detection of the perpendicularly polarised component of non-linear vacuum response, but even make the parallel polarised component experimentally accessible for the first time.
Remarkably, the angular separation of the signal from the intense x-ray probe enables precision measurements even in presence of pump fluctuations and alignment jitter.
This provides direct access to the low-energy constants governing light-by-light scattering.
\end{abstract}

\maketitle

%\tableofcontents

\paragraph{Introduction}

Vacuum fluctuations induce non-linear interactions of electromagnetic fields. This implies light-by-light scattering and violations of the superposition principle which predicts light rays to traverse each other without interacting. Within the Standard Model (SM) the leading effect is governed by quantum electrodynamics (QED), where a virtual electron-positron pair can couple electromagnetic fields \cite{Heisenberg:1936nmg,Weisskopf:1996bu}; see  \cite{King:2015tba,Karbstein:2019oej,Fedotov:2022ely} for reviews.

Macroscopic electromagnetic fields available in the laboratory fulfill $\{|\vec{E}|,c|\vec{B}|\}\ll E_{\rm S}$, with $E_{\rm S}=m^2c^3/(e\hbar)\simeq1.3\times10^{18}\,{\rm V}/{\rm m}$ set by QED parameters: the electron mass $m$ and elementary charge $e$.
If these fields vary on scales much larger than $\lambdabar_{\rm C}=\hbar/(mc)\simeq3.8\times10^{-13}\,{\rm m}$, their leading interactions are governed by ($c=\hbar=1$)~\cite{Euler:1935zz,Euler:1935qgl}
\begin{equation}
 {\cal L}_{\rm int}\simeq\frac{m^4}{1440\pi^2}\biggl[a\Bigl(\frac{\vec{B}^2-\vec{E}^2}{E_{\rm S}^2}\Bigr)^2 + b\Bigl(\frac{2\vec{B}\cdot\vec{E}}{E_{\rm S}^2}\Bigr)^2\biggr]\,. \label{eq:Lint}
\end{equation}
The constants $a$ and $b$ control the strength of the four-field couplings.
QED predicts these to have a series expansion in $\alpha=e^2/(4\pi)\simeq1/137$ and read \cite{Heisenberg:1936nmg,Ritus:1975pcc,Gies:2016yaa}
\begin{equation}
a=4\Bigl(1+\frac{40}{9}\frac{\alpha}{\pi}+\ldots\Bigr)\ \ \ \text{and}\ \ \
b=7\Bigl(1+\frac{1315}{252}\frac{\alpha}{\pi}+\ldots\Bigr)\,. \label{eq:a,b}
\end{equation}
The sub-leading terms $\sim\alpha$ result in corrections on the $1\%$ level. 

While there is evidence for light-by-light scattering in heavy-ion collisions \cite{dEnterria:2013zqi,Aaboud:2017bwk,Sirunyan:2018fhl,Aad:2019ock}, to date the interaction of macroscopic fields could not be detected in a controlled laboratory experiment.
A measurement of $a$ and $b$ would constitute a new precision test of QED and constrain the parameter space of BSM extensions such as Weakly-Interacting Slim Particles which are expected to leave an imprint on these. 
Corrections of the QED values by other SM sectors are suppressed with $(m/M)^4\ll1$ because their effective masses (charges) fulfill $M\gg m$ ($Q\sim e$).

A famous prediction of Eq.~\eqref{eq:Lint} is vacuum birefringence \cite{Toll:1952rq,Baier:1967zza}:
linearly polarised probe light traversing a pump field can obtain a $\perp$-polarised component and become elliptical. 
Progress in laser technology has resulted in realistic concepts to detect this effect in head-on laser pulse collisions \cite{Aleksandrov:1985,Heinzl:2006xc,King:2010kvw,Dinu:2013gaa,Karbstein:2015xra,Schlenvoigt:2016jrd,Karbstein:2016lby,Ataman:2018ucl,Shen:2018lbq,Tzenov:2019ovq,Ahmadiniaz:2020lbg} for the first time; see \cite{Iacopini:1979ci,Fan:2017fnd,Ejlli:2020yhk, Agil:2021fiq,Battesti:2018bgc} for searches in magnetic fields.
In this scenario, typically the number of $\perp$-polarised photons $N_\perp$, a subset of the total signal $N_\perp+N_\parallel$, constitutes the observable.
The scaling
\begin{equation}
 N_{\parallel,\perp}\sim c_{\parallel,\perp} \Bigl(\frac{I}{I_{\rm S}}\frac{\omega}{m}\Bigr)^2 N\,,
 \label{eq:scaling}
\end{equation}
with the pump intensity $I$ and the probe photon energy $\omega$ (number $N$) suggests the use of an XFEL as probe and a tightly focused high-intensity laser as pump for this experiment \cite{Heinzl:2006xc,DiPiazza:2006pr}; $I_{\rm S}=E_{\rm S}^2$.
The coefficients $c_\parallel=[a+b+(a-b)\cos(2\phi)]^2$ and $c_\perp=[(a-b)\sin(2\phi)]^2$ depend on $a$, $b$ and the relative polarisation $\phi$ of pump and probe; the ratio $c_\perp/c_\parallel$ depends on $\phi$ and $(a+b)/(a-b)$.
For collisions with a transverse impact parameter $r_0$ the signal photon numbers~\eqref{eq:scaling} decrease as \cite{Karbstein:2018omb}
\begin{equation}
 \sim\exp\biggl\{-4\Bigl(\frac{r_0}{w_0}\Bigr)^2\frac{1}{1+2(\frac{\mathfrak{w}_0}{w_0})^2}\biggr\}
 \label{eq:impact}
\end{equation}
from their maximum values at zero impact; $w_0$ ($\mathfrak{w}_0$) is the waist radius of the pump (probe) beam.

Detecting both $N_{\parallel,\perp}$ allows $a$ and $b$ to be inferred. As these depend on the driving fields in the same way, a simultaneous detection provides access to the ratio
\begin{equation}
 \frac{N_\perp}{N_\parallel}\Big|_{\phi=\frac{\pi}{4}} \simeq \Bigl(\frac{a-b}{a+b}\Bigr)^2=\frac{9}{121}\Bigl(1+\frac{260}{99}\frac{\alpha}{\pi}+\ldots\Bigr)\,. \label{eq:ratio}
\end{equation}
This observable does not depend on intensity and thus is insensitive to fluctuations in experimental parameters such as 
spatio-temporal jitter or intensity fluctuations. These typically limit the achievable precision in experiments requiring the overlap of pump and probe foci.

As the signal photons are predominantly emitted in the forward cone of the probe, discerning the signal from the background constitutes a formidable challenge.
Recent theoretical work showed that this is possible with a probe modified such as to exhibit a shadow in the far field while retaining a peaked focus profile \cite{Karbstein:2020gzg}.
Experimentally, this annular beam approach was pioneered by \cite{Peatross:1994,Hoerlein:2008} for the detection of weak non-linear optics signals in the presence of strong fields: blocking a part of the cross section of the original beam prior to focusing with a well-defined beam-stop creates a shadow in the collimated beam, which is then also present in the converging (expanding) beam before (after) focus. This arrangement can be seen as analogous to the commonly used spatial filtering techniques in linear optics.

\begin{figure}[htp]
    \centering
    \includegraphics[width=1\linewidth]{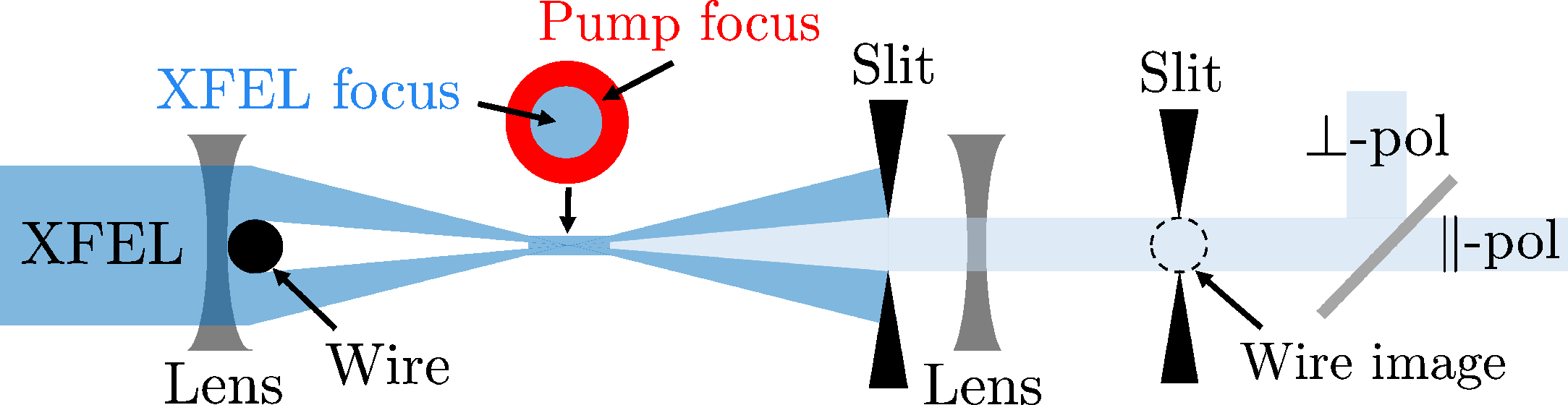}
    \caption{Schematic layout of an experiment to measure the coupling constants. The XFEL is focused to a spot with a wire creating a central shadow in the beam on both sides of the focus while retaining a central intensity peak in the focus.
    X-ray optics image the wire to a matched aperture plane. 
    The interaction with the pump results in signal photons scattered into the central shadow.
    The $\perp$, $\parallel$-polarised components are directed to separate detectors using a crystal polariser.}
    \label{fig:schema}
\end{figure}

In this letter, we combine the results of a first-principles calculation and a proof-of-concept experiment to show that this scheme is capable of determining both $a$ and $b$; see Fig.~\ref{fig:schema} for a schematic illustration.
Such a measurement could, e.g., be performed at the Helmholtz International Beamline for Extreme Fields (HIBEF) at the European XFEL  \cite{HIBEF}, SACLA \cite{SACLA} or LCLS \cite{LCLS}.

\paragraph{Theoretical scenario}
Both the pump and the counter-propagating x-ray probe are pulsed paraxial beams \cite{Karbstein:2019bhp}.
The waist $w_0=2\lambda f/\pi$ of the fundamental Gaussian pump (wavelength $\lambda$, pulse energy $W$, $1/{\rm e}^2$ duration $\tau$) is determined by the $f$-number of the focusing element.
The annular probe (duration $T$) is obtained by superposing two flattened Gaussian beams \cite{Gori:1994,Karbstein:2020gzg} characterised by far-field intensity profiles $I_{\cal N}(\vartheta)\sim[\Gamma({\cal N}+1,\vartheta^2/\theta^2)/{\cal N}!]^2$ with different integers ${\cal N}>{\cal N}'\geq1$;
the angle $\vartheta$ is measured from the beam axis, $\Gamma(\cdot,\cdot)$ is the incomplete $\Gamma$ function, and $\theta$ determines the beam divergence.
The outer (inner) radial divergence of this annular probe is $\theta_{\cal N}\approx\theta\sqrt{{\cal N}+1}$ ($\theta_{{\cal N}'}\approx\theta\sqrt{{\cal N}'+1}$), such that the ratio $(\theta_{{\cal N}'}/\theta_{\cal N})^2\approx({\cal N}'+1)/({\cal N}+1)$ measures the fraction of the cross section of the original probe ($N$ photons of energy $\omega$) to be blocked by the beam-stop.
Here we set $\theta=\frac{2}{\mathfrak{w}_0\omega}(\frac{2-2/{\rm e}}{{\cal N}+{\cal N}'+1+2/{\rm e}})^{1/2}$ ensuring the probe waist to be given by $\mathfrak{w}_0$ \cite{Karbstein:2020gzg}.
Upon this identification $I_{\cal N}(\vartheta)$ approaches a Heaviside function $\Theta(\theta_{\cal N}-\vartheta)$ for ${\cal N}\to\infty$.
Hence, for ${\cal N}\gtrsim{\cal N}'\gg1$, which we assume to hold in the remainder of this letter, the far-field intensity profile of the probe scales as $I^\circledcirc_{{\cal N},{\cal N}'}(\vartheta)\sim\Theta(\theta_{\cal N}-\vartheta)-\Theta(\theta_{{\cal N}'}-\vartheta)$.

A general expression for the far-field distribution ${\rm d}^3 N_{\parallel,\perp}/{\rm d}^3k$ of signal photons of wave vector $\vec{k}={\rm k}(\cos\varphi\sin\vartheta,\sin\varphi\sin\vartheta,\cos\vartheta)$ induced in the head-on collision of a fundamental Gaussian pump and a paraxial x-ray probe of generic mode composition was recently derived in \cite{Karbstein:2020gzg}; here the integration over the longitudinal coordinate $\rm z$ is still to be performed.
Note that in the conventions of this letter we have $r({\rm z})=\theta w({\rm z})\omega/2$ with pump radius $w({\rm z})$.
The result of \cite{Karbstein:2020gzg} serves as starting point for our calculation.
As $\theta\sim1/\sqrt{{\cal N}+{\cal N}'}\ll1$, functions of $r({\rm z})$ are slowly varying with $\rm z$ by definition.
Hence, the $\rm z$ integral can be evaluated with the saddle-point method; it can be easily checked that the leading order approximation captures the dominant contribution at ${\cal N}>{\cal N}'\gg1$.
Making use of the fact that the signal is predominantly emitted at ${\rm k}\simeq\omega$ and $\vartheta\ll1$, it moreover amounts to an excellent approximation to identify ${\rm k}=\omega$ in the overall prefactor and whenever $\rm k$ effectively occurs in combination with powers of $\vartheta$ \cite{Karbstein:2018omb,Karbstein:2019oej}.
Besides, we may neglect terms parametrically suppressed by $\omega\vartheta^2\ll1$.
Finally, we expand the individual factors constituting the signal photon emission amplitude in $\theta\ll1$ and keep only the leading contributions.
For $r_0=0$, this results in
\begin{align}
 &\frac{{\rm d}^3N_{\parallel,\perp}}{{\rm dk}\,{\rm d}\!\cos\vartheta\,{\rm d}\varphi}\simeq \frac{\alpha^4}{4050\pi^3}c_{\parallel,\perp}\frac{T}{2\sqrt{2\pi}}\,\theta^2\Bigl(\frac{\omega}{m}\Bigr)^6\Bigl(\frac{W}{m}\Bigr)^2 N \nonumber\\
 &\quad\quad\quad\quad\quad\quad\quad\times{\rm e}^{-(\frac{T}{2\sqrt{2}})^2({\rm k}-\omega)^2}\,\Bigl|\sum_{p=0}^{\cal N}{\cal S}_p\Bigr|^2\,,\nonumber \\
&\text{with}\quad\quad {\cal S}_p=
 \frac{c_{p,{\cal N}}-\Theta({\cal N}'-p)c_{p,{\cal N}'}}{\sqrt{2{\cal N}}}\nonumber\\
 &\quad\quad\quad\quad\quad\times {\rm e}^{-\frac{1}{2}(\frac{w_0\omega}{2})^2\vartheta^2} L_p\bigl(-\tfrac{1}{2}(\tfrac{w_0\omega}{2})^4\theta^2\vartheta^2\bigr)\,;\quad\quad \label{eq:diffsig}
\end{align}
$L_p(\cdot)$ are Laguerre polynomials and $c_{p,{\cal N}}=\sum_{k=p}^{\cal N}\binom{k}{p}\frac{1}{2^k}$.

Next, we aim at fixing $\mathfrak{w}_0$ such as to maximise the signal in the shadow for given $w_0$.
To this end, we demand the second derivative of Eq.~\eqref{eq:diffsig} for $\vartheta$ to vanish as this ensures the slowest drop of Eq.~\eqref{eq:diffsig} from its maximum at $\vartheta=0$ towards larger $\vartheta$.
This yields $\theta=\frac{2}{w_0\omega}(\frac{2}{{\cal N}+{\cal N}'+1})^{1/2}$, and thus $\mathfrak{w}_0\simeq w_0(\frac{1-1/{\rm e}}{2})^{1/2}\approx 0.6w_0$ for ${\cal N}\gtrsim{\cal N}'\gg1$.
The number of signal photons scattered into the shadow then scales as
\begin{equation}
 N_{\parallel,\perp}\sim \pi\theta_{{\cal N}'}^2\frac{{\rm d}^2N_{\parallel,\perp}}{{\rm d}\!\cos\vartheta\,{\rm d}\varphi}\Big|_{\vartheta=0} \sim \Bigl(\frac{{\cal N}-{\cal N}'}{{\cal N}+{\cal N}'}\Bigr)^2\frac{{\cal N}'}{\cal N}\,,
\end{equation}
which becomes maximum for ${\cal N}'=(\sqrt{5}-2){\cal N}\approx0.2{\cal N}$, i.e., when $(\theta_{{\cal N}'}/\theta_{\cal N})^2\approx20\%$ of the probe are blocked by the beam-stop.
Adopting this choice, we finally find that
\begin{align}
 N_{\parallel,\perp}&\approx \frac{16(5\sqrt{5}-11)\alpha^4 }{2025\pi^2}c_{\parallel,\perp}\Bigl(\frac{\omega}{m}\frac{W}{m}\Bigr)^2\Bigl(\frac{\lambdabar_{\rm C}}{w_0}\Bigr)^4 N \nonumber\\
 &\simeq 5.17\times10^{-18}\,c_{\parallel,\perp}\Bigl(\frac{\omega}{\rm keV}\frac{W}{\rm J}\Bigr)^2\Bigl(\frac{\si\micro{\rm m}}{w_0}\Bigr)^4 N \label{eq:result_Np}
\end{align}
signal photons are scattered in the central shadow in the beam, where the background is significantly reduced; the effect of a transverse impact $r_0\neq0$ can be estimated by multiplication with Eq.~\eqref{eq:impact}, which for the choices of parameters adopted here becomes $\approx\exp\{-2.3\,(r_0/w_0)^2\}$.
The strength of the remaining background component can be further reduced by imaging the beam-stop onto an aperture, allowing to eliminate the original beam fully.
The residual background is then due to scattering and diffraction.
For the experimental parameters $\lambda=800\,{\rm nm}$, $\omega=12.914\,{\rm keV}$ and $N=8.26\times10^{11}$ available at the European XFEL \cite{Schneidmiller:95609,Mosman:2021vua}, Eq.~\eqref{eq:result_Np} predicts $N_{\parallel,\perp}\approx 9.17\times10^{-3}f^{-4}\,c_{\parallel,\perp}(W/{\rm J})^2$.
For $\phi=\pi/4$, $W=10\,{\rm J}$ and $f=1$ this implies $N_\perp\approx8$ ($N_\parallel\approx107$) signal photons scattered into the shadow per shot.
While these photon numbers are easily detected by single photon sensitive detectors like CCDs, the challenge lies in designing the experiment such that any backgrounds are sufficiently suppressed. 
However, the very small wavelength and the small scattering cross-sections for crystalline media in the x-ray regime suggest the possibility of very low background count rates and hence high precision measurements. 
The availability of polarisers with differential transmission on the $10^{-11}$ level \cite{Schulze:2018,Schulze:2022} typically makes the weaker, $\perp$-polarised component easier to access.

To determine the feasibility of measuring both $a$ and $b$ we  demand the signal over $n$ shots $n N_{\parallel,\perp}$ (choose $n$) to be a factor \# above the standard deviation of the background $\sigma=\sqrt{nN_{\rm BG}}$, with $N_{\rm BG}$ background photons registered by the detector per shot, and calculate the required pump laser energy.  We concentrate on the $\parallel$-polarised component, as the added background rejection afforded by the use of polarising crystal easily outweighs the relatively small loss of signal. This results in the stringent criterion
\begin{equation}
    nN_\parallel>\#\sqrt{nN_{\rm BG}}\,, \label{eq:criterion}
\end{equation}
which for $\phi=\pi/4$ and the explicit experimental parameters given above yields a condition on the required number of shots $n$ for given values of $\#$, $N_{\rm BG}$, $W$ and $f$,
\begin{equation}
    n\gtrsim0.81f^8\,\#^2\frac{N_{\rm BG}}{(W/{\rm J})^4} \,. \label{eq:sigma}
\end{equation}
We emphasize the strong dependence on the $f$-number.

\paragraph{Measurement concept}
Most measurement concepts to date have focused on the $\perp$-polarised component, which allows rejection of the background using polarisers but is sensitive to alignment fluctuations and only provides access to $a-b$. The experimental concept envisioned here allows both $a$ and $b$ to be determined; see Fig.~\ref{fig:schema}. In this scenario the XFEL is focused and recollimated using a pair of compound refractive lenses (CRL). A  beam-stop (wire or disk) is placed in the beam centre and is imaged onto a matched aperture plane. An additional, intermediary aperture is placed before the collimating CRL to prevent scattering in that CRL.

As the direct path to the detector is geometrically blocked the background consists of XFEL photons  transmitted past the beam block and scattered in  the first CRL or diffracted from the beam block edge. The scattering in the CRL and can be estimated as  $N_1=P_{\rm s}\Omega N /(4\pi) \equiv \kappa N $, where $P_{\rm s} $ is the scattering  probability in the CRL and $\Omega$ the solid angle subtended by the slit/aperture. The contribution due to diffraction can be estimated using Fresnel diffraction theory \cite{BornWolf:1999}.
The primary source of suppression in the transmission $N_1$ is the small subtended angle of $\Omega/(4\pi)\sim\mathcal{O}(10^{-8})$ for a sub-mm aperture and $0.7\,{\rm m}$ focal length.
For high quality crystalline x-ray optics the diffuse  scattering probability is $\ll 10^{-3}$ \cite{Macrander:fv5142}.
Since the second set of CRL and slit acts identically to the first one the transmission due to scattered photons can be estimated as $N_2 = \kappa N_1 = \kappa^2N$. For the above parameters and a single XFEL pulse this value is well below the detection threshold.

\begin{figure}[htp]
    \centering
    \includegraphics[width=0.8\linewidth]{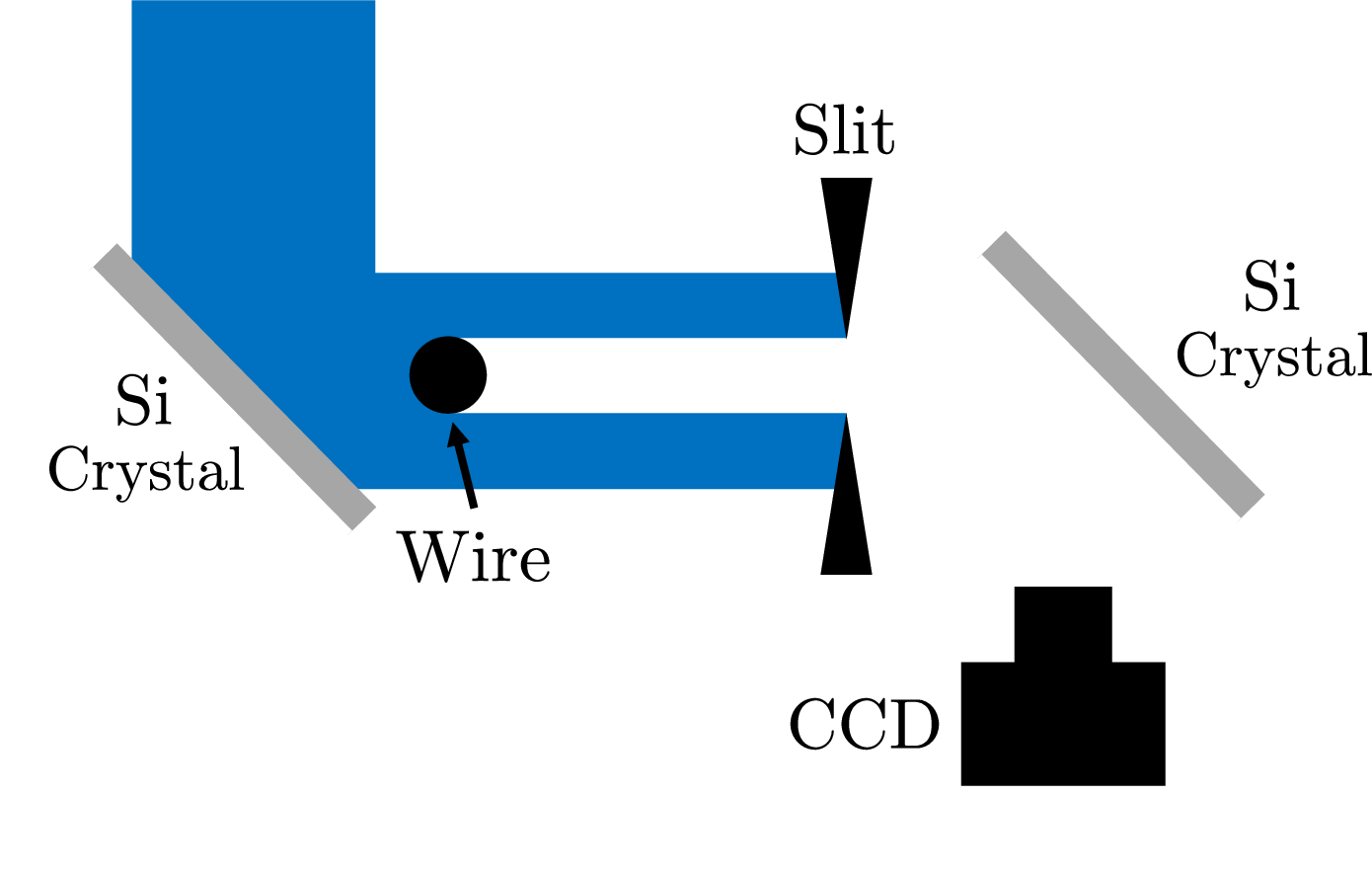}
        \includegraphics[width=1\linewidth]{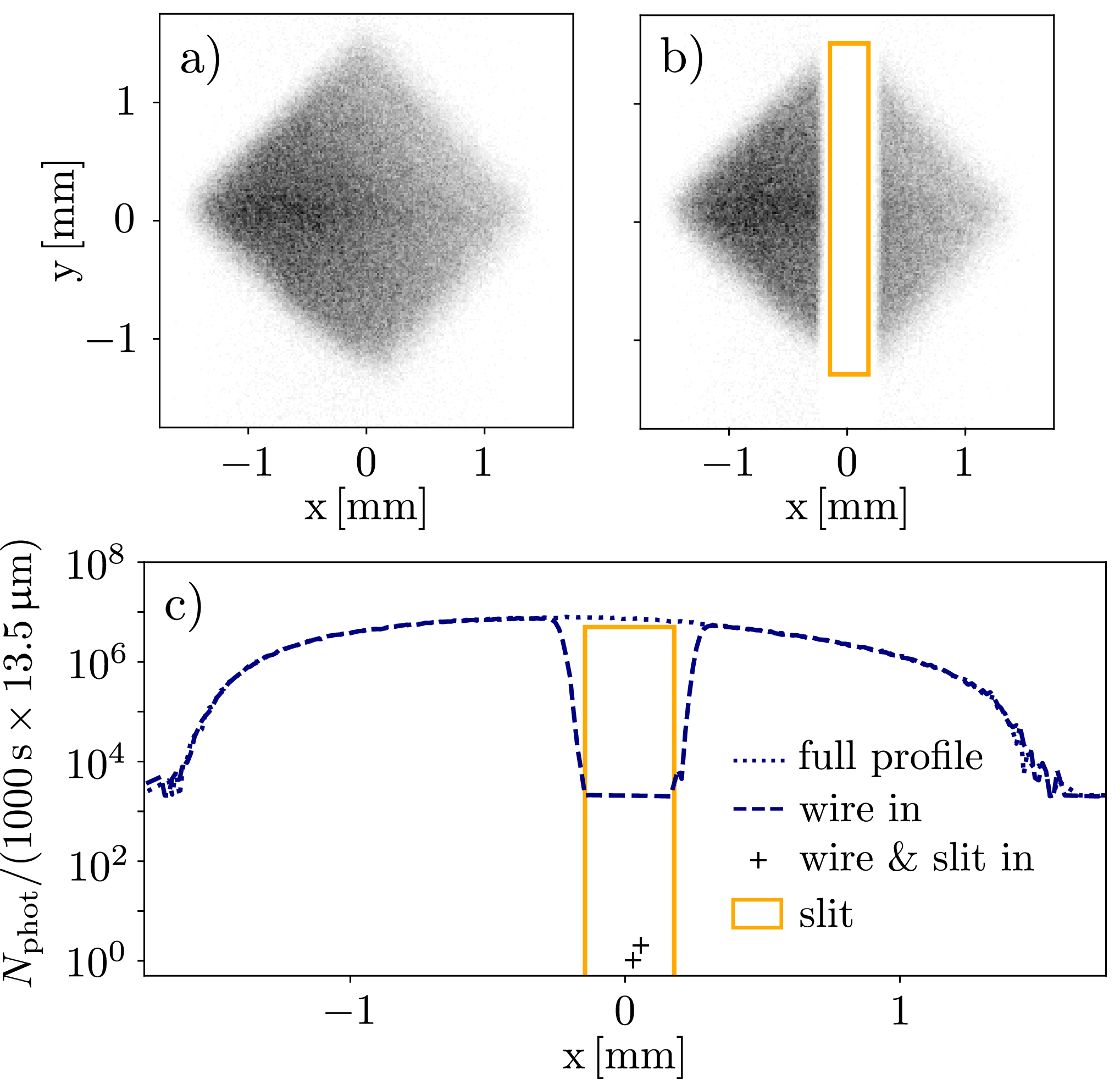}
    \caption{The top figure shows the simplified geometry used for the shadow contrast measurements. The middle figure displays $0.5\,{\rm s}$ exposures of the full x-ray beam profile a) without and b) with Au wire inserted. The green box indicates the slit position for shadow measurements. Figure c) depicts the number of photons per $1000\,{\rm s}$ integrated over the y-axis for different configurations.
    The $0.5\,{\rm s}$ exposures from a), b) are scaled to $1000\,{\rm s}$ resulting to a lower detection limit of  $N_{\rm phot}=2\times10^3$ in plot c). For high dynamic range measurements with a $1000\,{\rm s}$ exposure  a slit was used to block the transmitted beam on either side of the wire. With this arrangement three $8\,{\rm keV}$ photons were observed in the wire shadow consistent with level predicted due to diffraction and the off-Bragg-peak reflectivity of the silicon crystals.}
    \label{fig:measurement}
\end{figure}

The success of this scheme depends on the experimentally achievable purity of the shadow.  We therefore conducted a proof-of-concept experiment \cite{xray_shadow} at a $1.2\,{\rm  kW}$ rotating Cu-anode x-ray source to obtain an upper bound for the number of background photons expected to be scattered into the shadow.
Rather than using an annular structure we block the central portion of the x-ray probe with a thin Au wire of $500\,\upmu{\rm m}$ diameter. Our simplified setup shown schematically in Fig.~\ref{fig:measurement} (top) completely omits the lenses: the wire and the slit are placed in close proximity separated by only $4\,{\rm cm}$.
The beam is collimated by a confocal multilayer optic yielding $\approx5 \times 10^{9}$ Cu-K$\alpha$ photons per second with a symmetric divergence of $0.4\,{\rm mrad}$ within a rhombic beam profile with side lengths of $2\,{\rm mm}$.
A Si\,400 Bragg crystal is used as final scatterer before the wire, which also reduces the angular spread from $0.4\,{\rm mrad}$ to $18.8\,\upmu{\rm rad}$ determined by the rocking curve. 
A tungsten slit is placed approximately $4\,{\rm cm}$ downstream of the wire. For high sensitivity measurements the slit is adjusted to a width of $340\,\upmu{\rm m}$ to block the direct beam.
Scattering from the wire and slit is reduced by placing the assembly between two parallel, polished Si crystals using the 400 reflex. The signal is recorded with an x-ray CCD (Roper PI-MTE:2048B).

Figure~\ref{fig:measurement} shows the measured x-ray beam profile a) without and b) with the wire shadow for $0.5\,{\rm s}$ exposures. The wire introduces a shadow that eliminates all photons within the dynamic range of the camera for these short exposures.
Long ($1000\,{\rm s}$) exposures with the tungsten slit adjusted to block the direct beam are shown in Figure~\ref{fig:measurement} c) and compared to other configurations.  The angular acceptance of this geometry is $\Omega/(4\pi)=3\times10^{-5}$ and we observe a signal level of $(3\pm1.7)\times 10^{-8}$ of the primary beam intensity in the wire shadow. This value agrees well with the level of $2\times 10^{-8}$ predicted by Fresnel diffraction taking into account the  angular dependence of reflectivity in the second Si-crystal, which further suppresses the diffracted x-rays. The contribution due to scattering in the first Si-crystal is expected to be negligible in this geometry. A significant diffuse scattering contribution at twice the background level could be observed when an amorphous solid  ($135\,\upmu{\rm m}$ glass) was placed between the crystals. The observed result agrees very well with expectations and is close to the experimental sensitivity limit of our setup for reasonable integration times.

\paragraph{Implications for experiment}
As before we only consider the $\parallel$-polarised channel. To estimate the background suppression factor achievable in the setup with two lenses and slits with 1:1 imaging of the wire by the second CRL as shown in Fig.~\ref{fig:schema}, we estimate  scattering to be insignificant at $\kappa \approx 2\times 10^{-10}$ due to the smaller angle subtended and therefore $\kappa^2\approx 4\times10^{-20}$ in line with the results of our proof-of-concept experiment. As in the test setup the background levels will be dominated by diffraction. For a setup with CRLs with $0.7\,{\rm m}$ focal length double diffraction into the detector is estimated at the $2\times10^{-11}$ level with the signal in the $\perp$-channel further significantly suppressed by polarisation selection.

The predicted level of background suppression would allow for a single shot measurement of the observables for $W=10\,{\rm J}$ and $f=1$ or multi-shot measurements with relaxed laser requirements in terms of energy and/or focusing. As an experimental measurement of the background transmission level of the full setup  is beyond the measurement sensitivity of our test setup, we take  a conservative approach and assume that the full experiment with CRLs and two apertures would perform no better than our test setup at $3 \times 10^{-8}$, which implies $N_{\rm BG}=3 \times 10^{-8} N$.
Equation~\eqref{eq:sigma} predicts that even in this case a measurement of the $\parallel$-polarised signal component -- and thus also a direct determination of the fundamental constants~\eqref{eq:a,b} and a measurement of vacuum birefringence for the first time -- with a significance of $\#=5$ would require only $50$ shots for the above XFEL parameters and a laser energy of $W=10\,{\rm J}$ and $f=1$. Equation~\eqref{eq:criterion} predicts the required number of shots for a given value of $\#$ to scale as $n\sim(N_{\parallel,\perp})^{-2}$. Correspondingly, with the above experimental parameters and a repetition rate of $10\,{\rm Hz}$, in a one-day (one-year) measurement one could achieve a precision on the $10^{-3}$ ($10^{-5}$) level for $N_{\parallel,\perp}$, and thus on a corresponding level for a and b; cf. Eq.~\eqref{eq:criterion}. This implies the principle accessibility of higher-order corrections in $\alpha$ in Eq.~\eqref{eq:a,b}.
Finally, we note that the experimental scheme presented in this letter can be  extended to the optical regime where repetition rates of $100\,{\rm MHz}$ are feasible with cavities.
For sufficiently high circulating pulse energy this might facilitate quantum vacuum measurements with even higher precision.

\paragraph{Conclusions}
We have shown that the use of an XFEL probe modified  to exhibit a central shadow in both the converging and expanding beam makes the low-energy constants governing light-by-light scattering directly accessible experimentally for the first time.
These can be extracted from a simultaneous measurement of the $\parallel$ and $\perp$-polarised components of the non-linear response of the laser driven vacuum.
This constitutes a sensitive test of QED in an untested parameter regime and has the potential to probe and constrain physics beyond the Standard Model of particle physics.
Our approach also provides a realistic experimental platform for the first observation of vacuum birefringence in a controlled laboratory experiment: it has the potential to reduce the experimental time required for $5\sigma$ significance from previous estimates of about 6~days \cite{Schulze:2022} to less than one minute at currently available facilities such as HIBEF, and thus makes quantum vacuum non-linearities accessible with moderate experimental efforts for the first time.
Moreover, we emphasize that due to its high sensitivity our setup offers a great potential in searching for BSM extensions leaving an imprint on $a$ and $b$; see \cite{Fedotov:2022ely,Baker:2013zta} and references therein.

\acknowledgments

This work has been funded by the Deutsche Forschungsgemeinschaft (DFG) under Grant Nos. 416607684; 416702141; 416708866 within the Research Unit FOR2783/1.
We thank Ingo Uschmann for experimental support and acknowledge helpful discussions with IU and Tom Cowan.

\end{document}